# AI for All: Operationalising Diversity and Inclusion Requirements for AI Systems

Muneera Bano, Didar Zowghi, Vincenzo Gervasi, Rifat Shams

*Abstract*— **As Artificial Intelligence (AI) permeates many aspects of society, it brings numerous advantages while at the same time raising ethical concerns and potential risks, such as perpetuating inequalities through biased or discriminatory decision-making. To develop AI systems that cater for the needs of diverse users and uphold ethical values, it is essential to consider and integrate diversity and inclusion (D&I) principles throughout AI development and deployment. Requirements engineering (RE) is a fundamental process in developing software systems by eliciting and specifying relevant needs from diverse stakeholders. This research aims to address the lack of research and practice on how to elicit and capture D&I requirements for AI systems. We have conducted comprehensive data collection and synthesis from the literature review to extract requirements themes related to D&I in AI. We have proposed a tailored user story template to capture D&I requirements and conducted focus group exercises to use the themes and user story template in writing D&I requirements for two example AI systems. Additionally, we have investigated the capability of our solution by generating synthetic D&I requirements captured in user stories with the help of a Large Language Model.**

*Impact Statement* — **As AI systems become increasingly prevalent in everyday life, ensuring they respect and reflect the diversity of society is of paramount importance. Failing to address this issue can lead to AI solutions that perpetuate societal biases, inadvertently sidelining certain groups and prolonging existing inequalities. Our research proposes a mechanism for AI developers engaged in responsible AI engineering to seamlessly integrate diversity and inclusion principles during the system development, ensuring AI decisions and functionalities uphold ethical standards. Our proposal has the potential to directly influence further research on how AI systems are conceived, designed and developed, ensuring that they are inclusive of the needs of diverse users. On a social level, users can be more confident that advancements in AI technology would not come at the cost of marginalisation and potential discrimination.**

*Index Terms*— **Diversity, Inclusion, Requirements, Artificial Intelligence.**

## I. INTRODUCTION

The pervasive role of Artificial Intelligence (AI) in social interactions, from generating and recommending contents, to processing images and voices, brings numerous benefits but also necessitates addressing ethical implications and risks, such as ensuring equitable and non-discriminatory decision-making, and preventing the amplification of existing inequalities and biases [1]. Diversity and inclusion (D&I) in AI involves considering differences and underrepresented perspectives in AI development and deployment while addressing potential biases and promoting equitable outcomes for all concerned stakeholders [1]. Incorporating D&I principles in AI can enable technology to better respond to the needs of diverse users while upholding ethical values of fairness, transparency, and accountability [2].

Requirements engineering (RE) is well acknowledged to be an essential part of software development in general, and in developing AI systems in particular. RE includes the identification, analysis and specification of stakeholder needs, ideally captured in a consistent and precise manner. By focusing on users' and stakeholders' needs, RE aims to contribute to achieving user satisfaction and successful system adoption [3]. Using traditional RE practices for AI systems presents new challenges [4], as these traditional methods need to evolve to address new AI systems requirements, including those related to data and ethics [5, 6].

To develop ethical and trustworthy AI systems, it is recommended to consider embedding D&I principles throughout the entire development and deployment lifecycle [7]. Overlooking D&I aspects can result in issues related to fairness, trust, bias, and transparency, potentially leading to digital redlining, discrimination, and algorithmic oppression [8]. We posit that RE processes and practices can be tailored and adopted to identify and analyse AI risks and navigate trade-offs and conflicts that may arise due to neglecting D&I principles. For instance, RE can facilitate decision-making in scenarios where maximising inclusion may lead to reduced performance or efficiency or when data transparency about members of under-represented groups could compromise their privacy. By acknowledging diverse users and emphasising inclusive system development, RE can aid in achieving a balance between conflicting objectives and promote the development of ethical AI systems [9].

While a number of guidelines on ethical AI development exist [10], (e.g. addressing bias [11], fairness [12], transparency [13] and, explainable and responsible AI [14]), the published literature shows a scarcity of research about D&I in AI and to the best of our knowledge not much can be found on the topic from an RE perspective. Our research aims to fill this gap and to explore the operationalisation of D&I in AI guidelines into RE process.

Our research methodology encompasses three stages: 1) *data collection and analysis* from the published literature on D&I in AI to extract relevant themes, 2) proposing a *tailored user story template*, and 3) *focus group exercise* to explore the use of the extracted themes and user story template to specify D&I requirements for AI systems. Furthermore, given that involving many stakeholders with diverse attributes in requirements elicitation is challenging and time-consuming, we decided to explore the utility of Large Language Models in generating user stories from the D&I in AI themes. After each focus group exercise, we used GPT-4 to generate D&I user stories. We aimed to examine how closely the user stories from both human analysts and GPT-4 are aligned in terms of diversity attributes and covering the themes.

The major contributions of our work are:

- Addressing the research gap regarding elicitation of D&I requirements for AI systems.
- Identifying 23 unique D&I in AI themes from a comprehensive literature review.
- Introducing a tailored user story template for capturing D&I requirements for AI systems

The structure of this paper is as follows: Section II establishes a foundation by defining diversity and inclusion within the AI ecosystem. Section III presents the research motivation underpinning our work. Section IV details our research methodology, while Section V offers a summary of the results. Section VI delves into a discussion of the findings. Lastly, Section VII concludes the paper and proposes future research directions.

## II. DEFINING DIVERSITY AND INCLUSION IN AI

Despite the acknowledged importance of diversity and inclusion, there is a gap in the literature regarding how these principles can be practically implemented in AI systems. Fosch-Villaronga and Poulsen [15], define D&I in AI as a multi-faceted concept that addresses both AI's technical and socio-cultural aspects. They highlight diversity as the representation of individuals concerning socio-political power differentials such as gender and race. Inclusion, they suggest, is the representation of an individual user within a set of instances, with better alignment between a user and the options relevant to them, indicating greater inclusion. This concept is further analyzed at three levels: the technical, the community, and the user. The technical level considers whether algorithms account for all necessary variables and if they classify users in a discriminatory manner. The community level examines diversity and inclusivity in AI development teams, looking at gender representation and diversity of backgrounds. Finally, the user level focuses on the intended users of the system and how the research and implementation process takes into account the stakeholders and their feedback, emphasizing the principles of Responsible Research and Innovation.

The paucity of a comprehensive definition for D&I in AI within the existing literature has motivated us to propose a normative definition and a set of guidelines for ensuring these principles are incorporated into the AI development process. We have sought and received feedback iteratively on the definition and guidelines from Responsible AI and D&I experts [16]. We focused on a socio-technological perspective, recognizing that addressing bias and unfairness requires a holistic approach that considers cultural dynamics and norms and involves end users and other stakeholders. We defined D&I in AI as: *'inclusion' of humans with 'diverse' attributes and perspectives in the data, process, system, and governance of the AI ecosystem.* Diversity refers to the representation of the differences in attributes of humans in a group or society. Attributes are known facets of diversity, including (but not limited to) the protected attributes in Article 26 of the International Covenant on Civil and Political Rights (ICCPR), as well as race, color, sex, language, religion, national or social origin, property, birth or other status, and inter-sections of these attributes. Inclusion is the process of proactively involving and representing the most relevant humans with diverse attributes; those who are impacted by, and have an impact on, the AI ecosystem context.

We proposed that diversity and inclusion in AI can be structured and conceptualized involving five pillars: humans, data, process, system, and governance. The humans pillar focuses on the importance of including individuals with diverse attributes in all stages of AI development. The data pillar highlights the need to be mindful of potential biases in data collection and use. The process pillar emphasizes the need for diversity and inclusion considerations during the development, deployment, and evolution of AI systems. The system pillar recognizes the necessity for the AI system to be tested and monitored to ensure it does not promote non-inclusive behaviors. The governance pillar underlines the importance of structures and processes that ensure AI development is compliant with ethical principles, laws, and regulations. AI ecosystem refers to the 5 pillars (humans, data, process, system, and governance), plus the environment (i.e. application domain), within which the AI system is deployed and used.

## III. RESEARCH MOTIVATION

The increasing prominence of AI systems in our everyday lives has led to an urgent need for ethical and responsible AI consideration. Numerous ethical guidelines and principles for AI have emerged, emphasizing fairness, justice, and equity as crucial components of ethically sound AI systems [10, 17]. Despite the widespread recognition of D&I as essential social constructs for achieving unbiased outcomes [18, 19] , there is a glaring lack of concrete practical guidance on how to effectively integrate D&I principles into AI systems [1]. This gap has far-reaching implications for the broader AI ecosystem, as it may result in perpetuating existing biases, reinforcing social inequalities, and further marginalizing underrepresented groups. Inconsistencies in the interpretation and application of these principles [17], coupled with a lack of diversity in perspectives and underrepresentation of views from the Global South [7, 20], raise concerns about the effectiveness of current ethical AI guidelines.

Implementing ethical principles in AI remains challenging due to the absence of proven methods, common professional norms, and robust legal accountability mechanisms [21]. While AI ethics guidelines often focus on algorithmic decision-making, they tend to overlook the practical and operational aspects of the business practices and political economies surrounding AI systems [22]. This oversight can lead to issues such as "ethics washing", corporate secrecy, and competitive and speculative norms [21].

There is ongoing fierce debate on the effectiveness and practical applicability of AI ethical guidelines. Munn [23] highlights the ineffectiveness of current AI ethical principles in mitigating racial, social, and environmental harms associated with AI technologies, primarily due to their contested nature, isolation, and lack of enforceability. In response, Lundgren [24] argues that AI ethics should be viewed as trade-offs, enabling guidelines to provide action guidance through explicit choices and communication. Lundgren emphasizes operationalising ethical guidelines to make them accessible and actionable for non-experts, transforming complex social concepts into practical requirements. Despite conceptual disagreements, Lundgren suggests building on existing frameworks, focusing on areas of agreement, and setting clear requirements for data protection and fairness measures.

To address these challenges with an RE lens, and ensure AI systems are developed and deployed responsibly, we posit that it is crucial to operationalise diversity and inclusion requirements for AI. By providing clear, actionable steps for incorporating D&I principles in AI development and governance, we can promote a more inclusive, equitable, and ethical AI landscape [25-27]. We recognize the validity of Munn's critique, and in our research, we aim to bridge the gap between high-level ethical guidelines and practical implementation following Lundgren's call.

In detail, we study and explore the limited practical applicability of existing D&I guidelines in the context of RE for AI systems, emphasizing the need for operationalisation to make them effective. We found several issues with these guidelines, including their circularity, excess specificity for certain attributes and techniques, or conversely lack of sufficient specificity, and absolutism by ignoring resource constraints. While many guidelines are sensible as driving principles, we argue that transforming them into actionable requirements is crucial for integrating D&I concerns into AI system development. To address this challenge, we propose a user story template that maps the five pillars discussed in Section II onto *roles, artifacts,* and *processes,* aiming to make these guidelines more applicable in real-world software development contexts.

## IV. RESEARCH METHODOLOGY

As depicted in Figure 1, our research methodology consists of three stages: 1) Data Collection and Analysis from a Literature Review, 2) Tailoring a User Story Template for eliciting D&I in AI Requirements, and 3) Focus Group.

### A. Stage 1: Data Collection and Analysis

We built on our previous work, where data was gathered through two distinct approaches: a systematic literature review (SLR) [28], and a document analysis of published ethical AI guidelines related to D&I [16].

The SLR [28], among other objectives, aimed to identify challenges and their associated solutions (guidelines, strategies, approaches, or practices) associated with D&I in AI. After a rigorous search and selection, we found a sample of 48 academic papers from the period of 2017-2022. Open coding was applied to the data extracted from these papers, leading us to identify 45 challenges related to D&I in AI.

In our document analysis of grey literature, we applied a systematic approach to extract guidelines related to D&I in AI

from widely circulated sources, reports and grey literature [16]. We then conducted a thematic analysis of this list to identify 46 unique Guidelines about D&I in AI systems structured using the 5 pillars of D&I in AI definition.

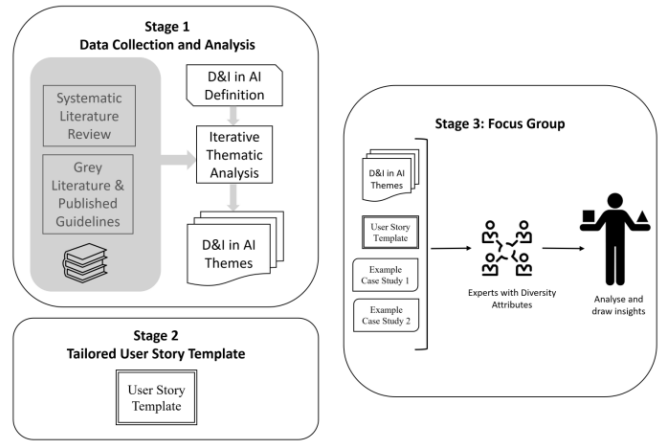

Fig 1. Research Methodology

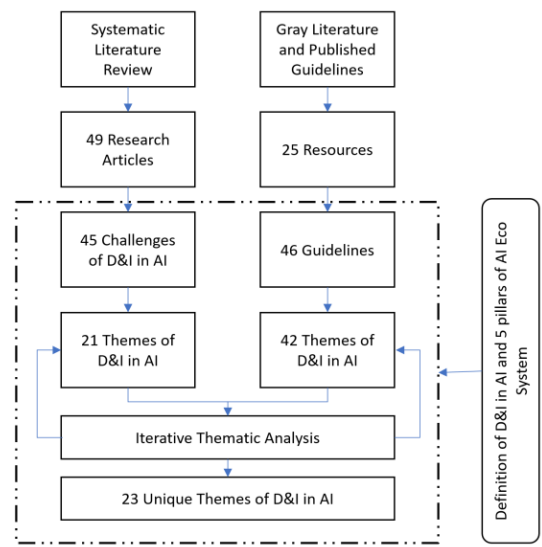

Fig 2. Data Analysis

The methodology depicted in Figure 2 outlines a rigorous and systematic approach we adopted to extract 23 unique themes related to D&I in AI from our two sources of data, SLR and Guidelines. Our complete dataset is provided as a spreadsheet[1]. The data from the literature review and guidelines contained a plethora of conceptual and semantic repetitions, necessitating meticulous refinement to extract only a unique and comprehensive list of themes. For the grey literature, the data collection and analysis were mainly undertaken by the second author, and the first author was involved in the analysis. Meanwhile, the SLR involved the collaborative efforts of the first, second, and fourth authors. These two sources individually revealed 45 challenges and 46 guidelines concerning D&I in AI, which after further analysis, were subsequently condensed into 21 and 42 distinct themes, respectively.

---

[1] Complete Dataset for Thematic Analysis:

https://docs.google.com/spreadsheets/d/1Zaqwa88nkePzy445nasEtXOzQreMsJDc/edit?usp=sharing&ouid=117004736404539149484&rtpof=true&sd=true

In the first level of thematic analysis, several themes were streamlined to eliminate semantic redundancies. For example, within the Human pillar of SLR, themes related to representation, including "imbalance of gender representation," "under-representation of marginalized groups," "lack of AI development team's diversity," and "lack of AI researcher's diversity," were consolidated under one theme of "Representation, Diversity and Inclusion". An iterative thematic analysis was applied, further refining and merging all the derived themes. This process resulted in identifying 23 unique themes of D&I in the AI.

### B. Stage 2: Tailored User Story Template

In the second phase, we designed a tailored User Story template for specifying D&I requirements in AI, that could be used for eliciting and capturing D&I in AI requirements by using the themes identified in stage 1. The template particularly focuses on roles that are embodied by persons with diverse attributes or that specify system behaviors that involve diverse attributes. Our hypothesis here is that a structured, specialized template can help human analysts focus specifically on D&I during requirements elicitation.

### C. Stage 3: Focus Group Exercise

Using the themes identified in stage 1 and the tailored user story template from stage 2, we presented the artifacts to a focus group of four experts (in RE, D&I and AI) along with two example case studies. They were asked to elicit D&I requirements for AI systems that were described in the case studies. The team of experts comprised of three females and one male participant, coming with different diversity attributes of age, race, faith, culture, language, and nationality.

### D. Exploring the utility of LLM

As we wanted to explore the utility of automating this process using LLM and examine how effective the process of writing user stories according to the template could be, we additionally used the popular LLM, GPT-4 from OpenAI as a tool to specify requirements. The test was based on the hypothesis that LLMs can support and complement humans in identifying relevant D&I in AI requirements and capture them in the user story template extracted from the template.

## V. RESULTS

### A. D&I in AI Themes

Table 1 provides a unique list of 23 themes extracted from our literature review. The themes are structured across five foundational pillars, showing the multifaceted dimensions of AI ecosystem.

The **'Humans'** pillar fundamentally addresses the integration of diverse human perspectives in AI. Within this pillar, themes such as #1 highlight the imperatives of ensuring representation, diversity, and inclusion throughout the AI lifecycle. It suggests that AI applications should be reflective of the diverse human experiences they are designed to serve. Complementing this, #2 underscores the pivotal role of stakeholders. It promotes the tenet that AI's potential is best harnessed when there is an active engagement and collaboration amongst all stakeholders, ensuring collective wisdom and diverse viewpoints shape AI's trajectory.

**Data** serves as the bedrock upon which AI operates. Within this pillar, theme #8 champions the ethos of transparency and explainability, signifying that the logic and reasoning behind AI's decisions should be open for scrutiny and comprehension. Similarly, theme #9 is related to data security, underscoring the principles of user privacy, data sovereignty, and infrastructure. These themes advocate for a data ecosystem that is both transparent and secure, upholding both trust and reliability.

| Pillar | # | Theme |
|---|---|---|
| **Humans** | 1 | **AI Lifecycle:** Representation, Diversity, and Inclusion |
| | 2 | **AI Stakeholder:** Engagement and Collaboration |
| | 3 | **AI Context:** Awareness and Conflict Management |
| | 4 | **AI Foundations:** Socio-technical Approach |
| | 5 | **AI Education:** Inclusive Infrastructure and Training |
| | 6 | **AI Opportunities:** Equitable Practices and Challenges |
| | 7 | **AI Challenges:** Inclusion Aspects |
| **Data** | 8 | **AI Data:** Transparency and Explainability |
| | 9 | **AI Data Security:** Privacy, Sovereignty, and Infrastructure |
| | 10 | **AI Data Modelling:** Selection and Development |
| | 11 | **AI Data Management:** Documentation and Examination |
| | 12 | **AI Data Analysis:** Bias and Inequity |
| | 13 | **AI Data Traits:** Demographic Considerations |
| **Process** | 14 | **AI Analysis:** Bias and Marginalization |
| | 15 | **AI Performance:** Evaluation, Monitoring, and Refinement |
| | 16 | **AI Design:** Trade-offs Considerations |
| **System** | 17 | **AI System Design:** Inclusive Design and Development |
| | 18 | **AI Awareness:** Bias Recognition and Understanding |
| | 19 | **AI Tools Evaluation:** Bias and Representation |
| | 20 | **AI System Usability:** Accessibility Assessment |
| **Governance** | 21 | **AI Strategy:** Policy and Governance |
| | 22 | **AI Safety Protocols:** Risk Management and Standards |
| | 23 | **AI Ethical Directives:** Equity, Diversity, and Inclusion Principles |

Within the **Process** pillar, theme #14 emphasising the importance of analyzing AI processes for inherent biases and potential marginalization. It reiterates that every step of AI's decision-making process should minimise unintended biases. Theme #15 relates to continuous refinement are instrumental to ensure AI's outputs are both effective and inclusive.

The **System** pillar relates to the various aspects of diversity and inclusion principles within design and development of AI systems. Theme #17 advocates placing inclusivity at the core of AI system development. This theme suggests that the architecture of AI systems should inherently cater to diverse user groups. Similarly, theme #18 highlights the importance of building AI systems with a clear awareness of biases, thereby reducing the chances of biased outcomes.

The **Governance** pillar provides overarching guidance for AI's ethical and operational conduct. Theme #21 highlights the need of a robust AI strategy, emphasizing the need for clarity in policy formulation and governance structures. This ensures that AI initiatives align with broader organizational or societal objectives. Complementing this, theme #22 focuses on the significance of safety protocols, underscoring the necessity for risk management and adherence to established standards, supporting the ethical principles within AI system development.

From our analysis, noteworthy insights emerge related to *conflict identification* and *trade-off analysis*. While analyzing diversity and inclusion in AI, balancing different themes can often necessitate conflict resolution. For example, consider theme #8 (AI Data: Transparency and Explainability) and theme #9 (AI Data Security: Privacy, Sovereignty, and Infrastructure). While transparency and explainability demand open and interpretable AI systems, prioritizing data security and user privacy might require certain data operations to remain unchanged.

Similarly, within the 'Humans' pillar, theme #2 (AI Stakeholder: Engagement and Collaboration) emphasizes the importance of a collective approach involving diverse stakeholders in AI development. However, as theme #3 (AI Context: Awareness and Conflict Management) suggests, such extensive collaboration can occasionally lead to conflicting viewpoints, especially where stakeholders come from diverse cultural or societal understandings. Prioritizing one theme over the other might be detrimental, so a trade-off analysis can help in striking a balance.

Similarly, within the 'System' pillar, theme #17 (AI System Design: Inclusive Design and Development) and theme #20 (AI System Usability: Accessibility Assessment) reveals another potential area for trade-off analysis. While an inclusive design ensures that AI systems are equitable and considerate of diverse users, an overemphasis on inclusivity might sometimes lead to overly complex systems. This complexity can impede the system's usability, especially for users who aren't technologically adept.

### B. User Story Template

As we need D&I requirements to be managed together with other requirements for the system, rather than being stated in isolation, they need to be expressed in a form that is amenable to effective requirements management. User stories [29, 30] are a convenient format for our purposes, as well as being widely known and applied in practice. We propose a user story template specifically tailored to D&I in AI systems. To this end, we first map the five pillars discussed in Section III onto a model of *roles*, *artifacts* and *processes* as shown in Figure 3.

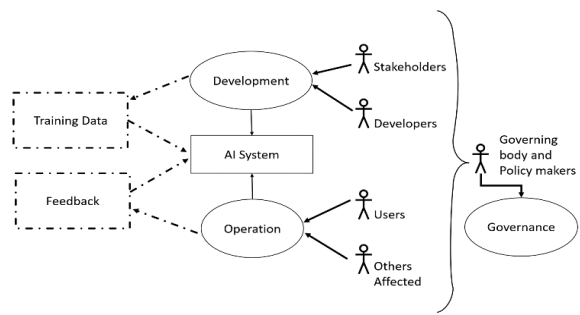

Fig 3. Elements of the User Story Template

We consider three artifacts, namely the AI system (corresponding to the "System" pillar), the initial training datasets and ongoing feedback if any (corresponding to the "Data" pillar), two processes of Development and Operation of the AI system (corresponding to the "Process" pillar), and an additional process, external to the development project in a strict sense, of Governance (corresponding to the "Governance" pillar). We include the establishment of laws, company policies, industrial standards, as well as education of the public at large and management of social expectations as part of the Governance process.

Finally, the Humans pillar is structured according to a variety of roles participating in the various processes; we single out *developers and other stakeholders* involved in the development process, *users* of the AI system and *others that are affected* by the operations or even mere existence of the AI system; as well as *governing body and policy makers*. All these roles should be interpreted with an expansive meaning, and with the understanding that a specific development project may well use more or less roles (or different naming). Similarly, a specific project may use different and more specific names (possibly of multiple specific sub-components) instead of "AI system", and similarly for the other elements in our depiction. The template we recommend to express D&I-related user stories is the following, with the specific rules listed below:

> As a *<role>*, I want *<role | process | artifact>* to *<predicate | behaviour>* so that *<rationale>*

This format helps to identify the key stakeholders, the desired action or outcome, and the underlying motivation behind the request. By employing this user story template, the development team can better address the needs of diverse stakeholders and align their AI systems with the overarching goals of diversity and inclusion.

To qualify as a proper D&I requirement, certain conditions on user stories must be fulfilled. **Firstly,** at least one instance of *role* should be qualified by a *protected attribute* and a value for that attribute (thus identifying a particular group of persons), or alternatively, the *predicate* or *behaviour* should refer to a protected attribute. This constraint ensures that the requirement represents a specific group's perspective or applies to that group, and hence may relate to a D&I issue. For example, a qualified role could be "a non-binary user of the system" (here the attribute is "gender identity" with value "non-binary") or "an Asian developer" (attribute "ethnicity" with value "Asian"). It must be stressed that the referenced group need not necessarily be a minority, underrepresented, or underprivileged group. The point of view of a majority group should also be listed among the requirements: inclusion is about including *all* points of view. An example of behavior that includes a protected attribute might be "provide audible feedback", with the rationale "so that visually impaired users are included".

The choice of which attributes and specific values are protected is often determined on a case-by-case basis by legal frameworks about non-discrimination, company policies, social expectations, or personal values.

**Secondly,** the *role*, *process* or *artifact* must be under the project owner's responsibility and control. This constraint ensures that only actionable guidelines are articulated as requirements. For instance, a small-scale AI project owner at a startup may not have the authority to dictate national government actions; thus, Theme 5 would not be amenable to be articulated as a requirement for their specific AI project. Another example is when developers have no control over the demographic makeup of anonymous feedback data, rendering a user story stating that feedback data must be demographically representative worthy, but not actionable. By adhering to these constraints, we may ensure that the generated user stories are indeed D&I requirements and can effectively guide the development of inclusive AI systems.

*C. Focus Group Exercise*

We conducted two focus group exercises with four experts who brought with them their diversity knowledge for gender, age, race, nationality, language, and religion. All of them are active researchers in the field of requirements engineering, ethical AI development, and diversity and inclusion. We presented them with two case studies: one on facial recognition and the other on voice recognition. In our effort to operationalize the D&I in AI themes, this approach allowed us to bring in practical scenarios, ensuring our guidelines were both theoretically sound and applicable to real-world AI challenges.

We recognised that engaging people with relevant diverse attributes and expertise for writing D&I in AI user stories is challenging and time-consuming. So, we decided to explore the utility of Large Language Models in generating user stories from the D&I in AI themes. After each focus group exercise, we provided the list of structured themes, user story template and example case studies, to GPT-4 and asked it to generate at least one user story per theme. We aimed to examine how closely the user stories from both human analysts and GPT-4 are aligned in terms of diversity attributes and covering the themes.

*1) Case Study 1*

In our first focus group, example case study 1 (Figure 4) was presented to four participants. The participants were provided with the themes and user story template and asked to write user stories.

Fig 4. Example Case Study 1

Following is the sample of the D&I requirements specified by the focus group, along with the diversity attribute that it highlights and the Theme from our Table 1 to which it aligns.

- *As a person with a visual impairment, I want FRASS to tell me the precise location where my face should be placed so that the Facial recognition system can scan my face properly for entry. [Disability, Theme 20]*

- *As an employee who smokes and has to take breaks for smoking outside throughout the day, I want the Work Hours Logging to recognise my needs to exit and re-enter the workplace for a smoking break. [Lifestyle, Theme 1]*

- *As a delivery person who has to deliver and move large boxes of equipment inside the building, I want the Real-time alert feature of the FRASS to recognise me as an authorised person to enter the premises. [Occupation, Theme 3]*

- *As a Muslim woman with niqab (covering face except eyes), I want FRASS to recognize me seeing my eyes only so that I don't have to open my face every time while accessing office. [Religion/Gender, Theme 1]*

- *As a person who wear glasses, I want FRASS to recognise my face accurately with it when scanning so that I don't have to remove it every time I enter or exit. [Physical Appearance, Theme 1]*

- *As an Asian lady with a black face, I want the FRASS system to correctly identify my face despite my skin tone so that it won't cause me to miss work or have a negative impact on my professional image. [Race, Theme 14]*

- *As a Manager who has to use reports generated by FRASS to assess performance of my team, I want FRASS to allow me to generate reports about the details of the movement of my team during working hours. [Occupation, Theme 21]*

- *As a transgender person who does not want to disclose my gender attribute, I want FRASS to respect my privacy. [Gender, Theme 1]*

- *As a developer of FRASS I want facial recognition feature to be inclusive as much as possible so that everyone in the company will build confidence on the system. [Occupation, Theme 1]*

- *As a person working in fire brigade with heavy uniform, I want FRASS to recognize me properly and quickly so that I can quickly manage my work hour logging and don't become late in any emergency situation at my workplace. [Occupation, Theme 3]*

- *As a female with a half-burnt face, I want the FRASS system to recognise my half-non-burnt face so that I do not need to uncover the burned area on my entry and exit every time. [Physical Appearance, Theme 6]*

- *As the CEO I want FRASS to generate overall reports on off-site hours of my employees so that I can take informed decisions. [Occupation, Theme 2]*

Here is the sample of the user stories written by GPT-4 for the example case study 1 and using the same themes and user story template.

- *As an employee of African descent, I want FRASS's facial recognition process to accurately identify diverse facial features so that I am not misidentified or denied access. [Race, Theme 1, 12, 18]*

- *As a female employee wearing a hijab, I want FRASS's facial recognition process to accurately recognize me even with my headgear on so that my religious beliefs do not hinder my ability to access my workplace. [Religion & Gender, Theme 1, 12, 18]*

- *As an employee with visual impairment, I want FRASS's access control mechanism to accommodate alternative identification methods so that I can enter the workplace without facing any barriers. [Disability, Theme 1, 20]*

- *As an elderly employee, I want FRASS's onboarding process to provide clear instructions and easy-to-understand tutorials so that I can comfortably and confidently use the system. [Age, Theme 5, 20]*

- *As a transgender employee, I want FRASS's data analytics process to respect my gender identity and not make assumptions based on facial features so that I feel represented and respected. [Gender Identity, Theme 1, 12, 18]*

- *As an employee who recently underwent facial surgery, I want FRASS's facial recognition system to easily allow for re-*

registration of my facial data so that I don't face difficulties accessing the premises after my recovery. [Physical Appearance, Theme 10, 15]

- As a young Asian female employee, I want FRASS's feedback and improvement process to actively seek diverse perspectives, including mine so that the system continuously evolves to cater to a broad range of users. [Race & Gender & Age, Theme 1, 2, 15]

The D&I requirements captured by the focus group of human analysts highlight a broader set of diversity attributes, considering aspects such as lifestyle, occupation, and different physical appearances beyond facial features. For instance, the requirements address employees with specific needs like smoking breaks and those professions, such as fire brigade personnel or delivery persons. Comparatively, GPT-4's generated requirements primarily gravitated towards more commonly discussed diversity attributes such as race, gender, religion, age, physical appearance, and disability. While both sets touch upon religion, physical appearance, race, and gender, the human-specified requirements offer a richer variety in context and rationale. The overlap in terms of disability, race, gender identity, and religion suggest that both the human group and GPT-4 acknowledge these as significant considerations. However, the human-generated requirements seem to demonstrate a deeper understanding of nuanced everyday experiences and challenges faced by diverse individuals in a workplace setting.

*2) Case Study 2*

We conducted our second focus group exercise (with the same participants) this time with example case study 2 (figure 5). Following is the sample of the D&I requirements specified by the focus group:

Fig 5. Example Case Study 2

- As a female health professional I want VRIMA to recognise my spoken words accurately and follow the instructions I give using my voice. [ Gender, Profession, Theme 17]
- As a covid positive doctor, I want VRIMA to recognize my voice even it is completely broken, so that I can conduct my regular task. [Health condition, Profession, Theme 20]
- As a doctor with a strong Persian accent I want VRIMA to recognise my ethnicity and hence understand my accent when I give instructions. [ Ethnicity, Accent, Profession, Theme 13]

- As a female health professional with an Asian accent, I want VRIMA to correctly recognise my voice to generate the correct health results in the form of images. [Gender, Profession, Accent, Ethnicity, Theme 13]
- As a health professional with a stammer, I want VRIMA to recognize my instruction and follow them accurately, so that my stammer does not result in incorrect outcomes. [ Health condition, Profession, Theme 20]
- As a health professional, I need to visit different wards in hospital which is noisy. I want VRIMA to suppress the background noises when I talk, so that it does not misunderstand what I am asking it to do. [Profession, Environment, Theme 18]
- As a health worker with speech disability, I want VRIMA to be trained with my impairment so that it can understand my instructions and capture it accurately. [Health condition, Profession, Theme 20]
- As a soft spoken doctor, I want VRIMA to identify my commands, so that I don't want to put extra effort while using it during my day to day activities. [Personality trait, Profession, Theme 1]
- As a person with visual impairment who cannot read text on screen, I want VRIMA to read back to me what has been captured as my instructions so that I can verify it. [Disability, Theme 20]
- As a Mom of a 2 years old child and being a health professional, I want VRIMA to differentiate my baby's not meaningful words and my commands, so that it doesn't misunderstand my command. [Parenthood, Profession, Theme 4]
- As a doctor who treats Sri Lankan patients with long names, I want VRIMA to capture the full name of the patients with a couple of parts of the names I tell to it, so that I do not want to tell the whole name every time. [Profession, Ethnicity, Theme 13]
- As a health care manager with slow speech, I want VRIMA to recognise my voice commands accurately, govern the performance of the newly installed AI system at the clinic, and generate results in the form of reports. [Profession, Personality trait, Theme 21]

Here is the sample of the user stories written by GPT-4 for the example case study 2 and using the same themes and user story template.

- As a non-native English-speaking healthcare professional, I want VRIMA to accurately recognize and transcribe my accent and dialect so that my patient records are correctly documented without any misunderstandings or biases. [Language, Accent, Theme 13]
- As a healthcare professional attending to diverse patients, I want VRIMA to provide culturally appropriate voice-guided instructions during medical procedures so that patients feel respected and at ease. [Culture, Ethnicity, Theme 3]
- As a healthcare professional, I want VRIMA to recognize and address non-binary and transgender patients correctly, providing instructions or feedback that's sensitive to their gender identity so that the care I provide is inclusive and respectful. [Gender, Sex, Theme 7]
- As a healthcare professional with a hearing impairment, I want VRIMA to offer visual cues or transcriptions in addition to voice prompts so that I can effectively utilize the system during my procedures and tasks. [Disability, Theme 20]
- As a healthcare professional treating patients with diverse backgrounds, I want VRIMA to acknowledge and provide insights on diverse medical histories and practices (e.g., traditional medicines) so that I can offer a more holistic approach to patient care. [Culture, Ethnicity, Theme 1]
- As a healthcare professional, I want VRIMA to alert me about potential drug interactions or issues that may conflict with a patient's religious or cultural beliefs so that the treatment plan is respectful and in alignment with their beliefs. [Religion, Culture, Theme 1, 23]

In a comparative analysis of D&I requirements elicited from a focus group and those generated by GPT-4, several distinctions emerge. The focus group's user stories seem to be rooted in tangible experiences, emphasizing specific challenges such as voice recognition across various genders, accents, and health conditions affecting speech. Notably, their concerns also included environmental challenges like noisy hospital settings and the confluence of personal and professional demands. In contrast, GPT-4 provided a more encompassing perspective, including wider diversity themes such as cultural and religious sensitivities, and a broader representation of gender and sex diversity. While the focus group offered detailed, practical insights, GPT-4 presented a more generalized, but inclusively nuanced perspective. Together, both datasets cover a comprehensive understanding of the multifaceted D&I needs in healthcare settings.

## VI. DISCUSSION

In this section we briefly discuss a few insights that are relevant in the wider context in which our proposed process for deriving D&I-focused user stories should take place, as well as discussing limitations of our research.

### A. Importance of Longterm Trade-off Analysis

In the competitive AI-driven business landscape, prioritizing D&I over accuracy and profit may seem like a hard sell. However, it is essential to recognize that neglecting D&I can have adverse long-term effects on a company's reputation, including backlash, loss of user trust, and potential legal consequences. Recognizing that businesses may initially benefit from the higher accuracy and efficiency of AI systems that do not fully address D&I concerns, and may be tempted to save on related costs, the approach we have presented in this paper helps focus the investment on those facets of D&I that really matter in the specific context, by way of considering among the requirements only the user stories focusing on the attributes, values, roles and artifact that are relevant for the project.

### B. Responsibility of the AI development team

While biased training datasets can perpetuate and reinforce discriminatory algorithms due to the use of shared and reusable datasets containing inherent biases [11] [31], a lack of diversity and inclusion in AI development teams can lead to unconscious biases and stereotypes, as team members may inadvertently project their own perceptions of reality or society into their work [32, 33].

### C. D&I Requirements and Non-Functional Requirements

Analysing D&I requirements in AI, like other non-functional requirements, entices concepts such as satisfaction up to a certain level (*satisficing*), inter-requirements structures based on positive and negative contributions to certain goals, and the determination of acceptable trade-offs in case of conflicts. By embodying general D&I guidelines into concrete user stories, our approach empowers developers to give them the same level of attention and rigour as other vital system characteristics. This approach ensures that AI systems are not only efficient and reliable but also ethically responsible and inclusive, catering to diverse user needs.

### D. Context and Culture

The context surrounding AI system development plays a crucial role in implementing diversity and inclusion requirements. For example, Europe's General Data Protection Regulation (GDPR) [34] enforces strict privacy laws that organizations must adhere to when designing AI systems; this in turn may affect whether certain D&I requirements can be satisfied. In contrast, some countries with less individualistic cultures may have more hands-off privacy norms. The realization of D&I requirements in AI systems is also influenced by various factors such as organizational culture, governmental and legal frameworks, and societal norms that govern the development and deployment of these systems.

### E. Using GPT-4 for writing D&I in AI Requirements

GPT-4, with its expansive training dataset, is well-equipped to assume diverse personas and extrapolate a broad range of requirements, reflecting potential challenges from myriad perspectives. This capacity aids in generating a holistic view of potential requirements, especially in the early stages of AI system design. However, while it can simulate multiple viewpoints, it inherently lacks the authenticity and depth of real human experiences, especially regarding nuanced aspects of diversity and inclusion. The subjective intricacies and tangible challenges faced by individuals in their lived experiences often provide invaluable insights that may not be entirely understood by GPT-4. Thus, combining the broad, simulated perspectives of GPT-4 with the grounded, real-world experiences of human stakeholders can offer a synergistic approach. Together, they provide a richer, more comprehensive information base that can immensely benefit analysts and developers, ensuring AI systems are both inclusive and cognizant of the intricate facets of human diversity.

### F. Limitations

Our research, despite employing a rigorous evidence-based methodology to identify D&I requirement themes, has certain limitations in coverage and applicability. One limitation is that the themes are derived solely from published and publicly available research, which might not cover the entire spectrum of D&I issues in AI systems. Additionally, although the identified requirements are generalizable and applicable to a broad range of AI systems, further analysis is necessary to tailor and adapt them to the specific context of an individual AI system project. The process of deriving D&I user stories that we have outlined is highly reliant not only on the requirements engineer's professional ability but also on their ethical and social sensibility. We believe this to be necessary and the only way to give due consideration to the nuances of ethical reasoning.

These limitations imply that the current research serves as a foundation for understanding D&I requirements in AI, but there is a need for more in-depth examination and customization to address unique challenges and peculiarities that may arise in specific AI implementations and social contexts.

## VII. Conclusion and Future Work

In this paper we have presented our extensive research on exploring D&I in AI guidelines and our attempt to operationalise them in the process of specifying D&I in AI requirements. We have identified 23 unique themes related to D&I in AI considerations from our literature review. We introduced a user story template for articulating D&I in AI requirements, and we have conducted a focus group with four human analysts to develop user stories for two cases of AI systems to gain insights into the process of writing D&I in AI requirements. Furthermore, we have explored the utility and usefulness of using GPT-4 as an agent in the automation of generating D&I in AI requirements. Comparing the user stories developed by human analysts and those generated by GPT-4 we have gained insights into the pros and cons of the use of LLM in this activity and the complementary nature of this form of human-machine collaboration. There remains a need for further exploration of the influence of cultural and legal contexts on the implementation of diversity and inclusion requirements in AI. Investigations could delve into how variances in privacy laws, data protection regulations, and cultural perspectives impact AI system development across different regions. Moreover, research could be undertaken to assess the effects of various AI systems, such as recognition technology, on individuals from diverse backgrounds encompassing race, gender, and age.

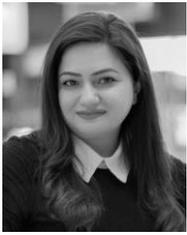

**1st Author: Muneera Bano, PhD** is Senior Research Scientist and member of Diversity and Inclusion team at CSIRO's Data61. She is an award-winning scholar, is passionate advocate for gender equity in STEM. She is a Diversity Inclusion and Belongingness (DIB) officer at Data61 and a member of the 'Equity, Diversity and Inclusion' committee for Science and Technology Australia. Muneera graduated with a PhD in Software Engineering from UTS in 2015. She has published more than 50 research articles in notable international forums on Software Engineering. Her research, influenced by her interest in AI and Diversity and Inclusion, emphasizes human-centric technologies.

Contact her at muneera.bano@csiro.au
Official Webpage: https://people.csiro.au/B/M/muneera-bano

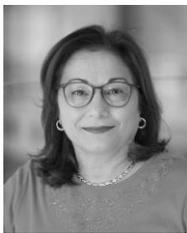

**2nd Author: Didar Zowghi,** (PhD, IEEE Member since 1995) is a Senior Principal Research Scientist and leads the science team for Diversity and Inc(lusion in AI at CSIRO's Data61. She is an Emeritus Professor at the University of Technology Sydney (UTS) and conjoint professor at the University of New South Wales (UNSW). She has decades of experience in Requirements Engineering research and practice. In 2019 she received the IEEE Lifetime Service Award for her contributions to the RE research community, and in 2022 the Distinguished Educator Award from IEEE Computer Society TCSE. She has published over 220 research articles in prestigious conferences and journals and has co-authored papers with over 100 researchers from 30+ countries.

Contact her at didar.zowghi@csiro.au
Official Webpage: https://people.csiro.au/z/D/Didar-Zowghi

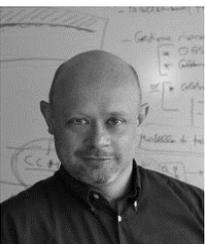

**3rd Author: Vincenzo Gervasi, PhD** is an associate professor in the University of Pisa's Computer Science Department. His research focuses on natural language processing applied to requirements engineering, formal specifications, and software architectures, fields in which he has published over 140 papers in international venues. Prof. Gervasi received his PhD in computer science from the University of Pisa and is a member of IFIP WG 2.9.

Contact him at vincenzo.gervasi@unipi.it

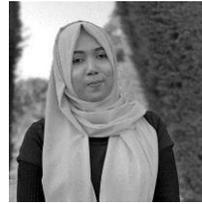

**4th Author: Rifat Ara Shams, PhD,** is a Postdoctoral Fellow in the AI Diversity and Inclusion team at CSIRO's Data61. Her research interests include Diversity and Inclusion in AI, Software Engineering for Responsible AI, Human Values in Software Engineering, Values and Ethics in AI. She completed her PhD in Software Engineering from Monash University, Australia in 2022.

Contact her at rifat.shams@csiro.au
Homepage: https://sites.google.com/view/rifatarashams.